\documentclass[%
reprint,
superscriptaddress,
 amsmath,amssymb,
prl,
longbibliography
]{revtex4-2}

\usepackage{ulem}
\usepackage{graphicx}
\usepackage{color} 
\usepackage{dcolumn}
\usepackage{bm}
\usepackage{amsmath}
\usepackage{dashrule}

\begin{document}

\title{Dynamical approach to the jamming problem}

\author{Sam Wilken}
\thanks{These two authors contributed equally}
\affiliation{%
Materials Department, University of California, Santa Barbara, California 93106, USA
}%
\author{Ashley Z. Guo}
\thanks{These two authors contributed equally}
\affiliation{%
Center for Soft Matter Research, Department of Physics, New York University, New York, NY 10003 USA
}%
\author{Dov Levine}
\affiliation{%
Department of Physics, Technion-IIT, Haifa, 32000 Israel
}%
\author{Paul M. Chaikin}
\affiliation{%
Center for Soft Matter Research, Department of Physics, New York University, New York, NY 10003 USA
}%

\begin{abstract} 

A simple dynamical model, Biased Random Organization, BRO, appears to produce configurations known as Random Close Packing (RCP) as BRO's densest critical point in dimension $d=3$. 
We conjecture that BRO likewise produces RCP in any dimension; if so, then RCP does not exist in $d=1-2$ (where BRO dynamics lead to crystalline order).
In $d=3-5$, BRO produces isostatic configurations and previously estimated RCP volume fractions 0.64, 0.46, and 0.30, respectively. 
For all investigated dimensions ($d=2-5$), we find that BRO belongs to the Manna universality class of dynamical phase transitions by measuring critical exponents associated with the steady-state activity and the long-range density fluctuations.
Additionally, BRO's distribution of near-contacts (gaps) displays behavior consistent with the infinite-dimensional theoretical treatment of RCP when $d \ge 4$.
The association of BRO's densest critical configurations with Random Close Packing implies that RCP's upper-critical dimension 
is consistent with the Manna class $d_{uc} = 4$.

\end{abstract}

\maketitle  

In the lore of sphere packings, if one throws identical ball bearings randomly into a box, they will invariably come to a dense disordered arrangement, conventionally called Random Close Packing (RCP), with volume fraction  $\phi_{RCP} \approx 0.64$.
By contrast, spheres placed in an ordered FCC lattice will fill space to $\phi_{FCC} = \frac{\pi}{\sqrt{18}} \approx 0.74$.  
The densest packings have $\phi_{max} = \phi_{FCC}$, which has been proven mathematically~\cite{hales2017formal}, but RCP, which might be thought of as the highest density ``random state", remains ill-defined~\cite{torquato2000random}.
Despite this, RCP has been used to gain insight into complex problems such as
the structure of liquids and glasses, the crystal-liquid transition, and the physics of jamming, where RCP corresponds to the $T=0$ critical point $\phi_J$ of the jamming transition~\cite{liu1998jamming,o2002random}.

There is general agreement that RCP can be constructed both  experimentally~\cite{scott1960packing,bernal1960packing,nowak1998density} and via simulation~\cite{lubachevsky1990geometric,lubachevsky1991disks,zinchenko1994algorithm,o2003jamming}, and generally possesses structural characteristics like isotropy, isostatic coordination, power-law decay of the nearest neighbor gap distribution~\cite{charbonneau2012universal} and hyperuniform density fluctuations (i.e. suppressed on long length scales)~\cite{donev2005unexpected,Wilken2020RCP}.
We previously found that, in $d=3$, the densest critical configurations of a Manna class model called Biased Random Organization (BRO) displayed many of the structural characteristics of Random Close Packing (e.g. volume fraction, isostatic coordination, hyperuniformity)~\cite{Wilken2020RCP}.
The Manna (or Conserved Directed Percolation) universality class contains models that undergo a dynamical second-order phase transition between quiescent (absorbing) and continuously evolving (active) states which are characterized by critical exponents that determine the dynamical and structural properties on both sides of the transition and an upper-critical dimension $d_{uc}$, above which critical behavior scales with mean-field exponents~\cite{LUBECK2004,henkel2008non}.

\begin{figure}
  \centering
  \includegraphics[width=.87\linewidth]{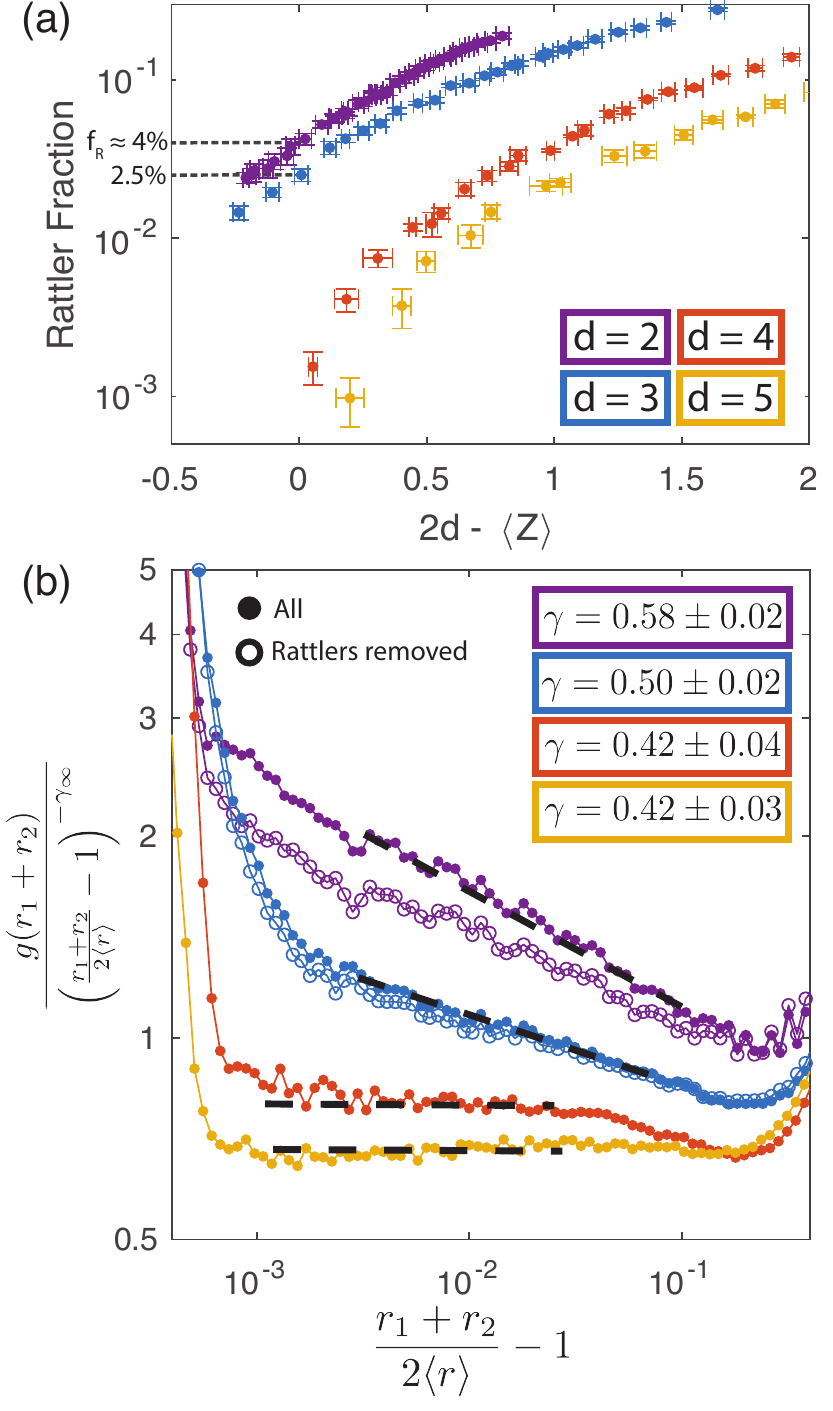}
  \caption{\textbf{Near-contact gap distribution and rattlers.} 
  (a) Fraction of rattlers ($Z<d+1$) approaching the critical point ($Z = 2d$) from the absorbing side ($\phi < \phi_c$). 
  (b) Distribution of gaps between neighbors (i.e. the pair correlation function $g(r_1 + r_2)$) for critical configurations near RCP in $d=2-5$ considering all particles (closed circles) and rattlers removed (open circles).
  Power-law scaling of near contacts (dashed lines), $g(r_1 + r_2) \sim ((r_1 + r_2)/2\langle r \rangle - 1)^{-\gamma}$, agrees with the infinite dimensional case ($\gamma_{\infty} = 0.413$) {\it only} for $d=4, 5$. 
  Removing rattlers decreases $\gamma$ slightly ($\gamma_{d=2} = 0.054 \pm 0.03$, $\gamma_{d=3} = 0.49 \pm 0.02$) but rattlers do not account for the deviation from the infinite dimensional value in $d=2,3$. 
  Curves are normalized by infinite dimensional scaling and shifted vertically for clarity.
  All simulations run with $N=27,000$ particles and $\epsilon/r = 0.001$.
  }
  \label{Gaps}
\end{figure}

In what follows, we conjecture that the densest BRO configurations correspond to RCP ensembles in all dimensions, and examine the consequences for RCP.  
We note here two findings that are at odds with other literature: (1) We find that there are no RCP configurations for mondisperse disks in $d=2$~\cite{atkinson2014existence,Zaccone2022}, and  (2) We observe RCP behavior with an upper-critical dimension $d_{uc} = 4$ rather than $2$~\cite{goodrich2012finite,hexner2019can,charbonneau2021finite,o2003jamming}.
Specifically, we show in this Letter that:
\begin{itemize}
 \item BRO remains in the Manna universality class on approach to RCP densities.
\item At or above the Manna class upper-critical dimension ($d\ge d_{uc} = 4$), BRO number fluctuations are random (Poisson) $\langle N^2 \rangle - \langle N \rangle^2 \sim \langle N \rangle \sim R^d$, where R is the linear size of the region sampled.
    \item The densest critical packings of BRO are isostatic, as expected for RCP.
    \item The gap distribution exponent $\gamma$~\cite{doi:10.1146/annurev-conmatphys-031016-025334} for critical BRO states agree with the infinite dimensional value $\gamma_\infty = 0.41269$ {\it only} at or above $d_{uc} = 4$.
    \item Removing rattlers, particles that are not rigidly constrained by the packing ($Z<d+1$), does not significantly impact $\gamma$ (Fig.~\ref{Gaps}b).
   
\end{itemize}
These observations lead us to propose that, in arbitrary dimension, RCP states may be given precise definition as the densest critical configurations of BRO.  
Further, since the Manna universality class has upper critical dimension $d_{uc}=4$~\cite{henkel2008non}, we predict that critical exponents for RCP in $d\ge4$ will be the same, and will coincide with the infinite dimensional case.

\begin{figure}
  \centering
  \includegraphics[width=\linewidth]{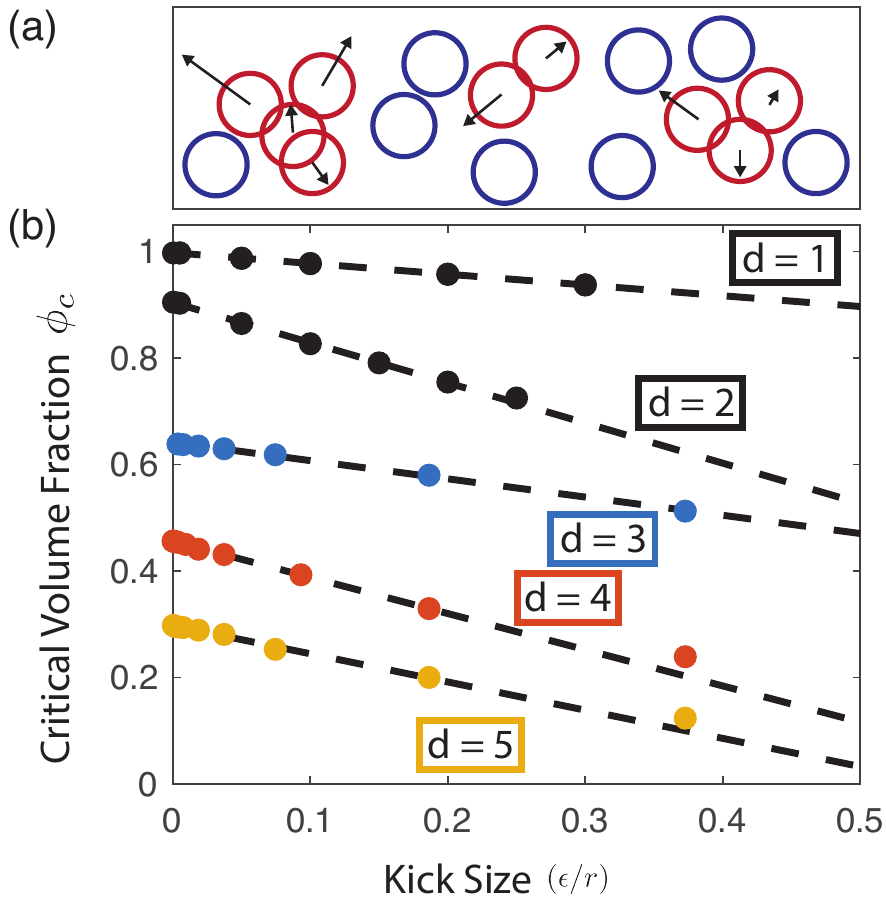}
  \caption{(a) Biased Random Organization (BRO) dynamics: particles receive displacements (kicks) away from overlapping (red) neighbors with random magnitude in [0,$\epsilon$]. Isolated particles (blue) do not move. (b) Critical densities, $\phi_c(\epsilon)$, of monodisperse packings in various dimensions as a function of $\epsilon$, in units of particle radius $r$. The densest critical structures, found via linear extrapolation (dashed lines), are {\it crystalline} in $d=1,2$ with $\phi_c(\epsilon \rightarrow 0) \approx 1.000, \, 0.905$. For $d=3,4,5$, they are {\it disordered}: $\phi_c(\epsilon \rightarrow 0) \approx \, 0.640, \, 0.46$ and $0.30$.  }
  \label{critical densities}
 \end{figure}

The BRO model we study is a variant of the Random Organization (RO) model, originally developed to study the structure and dynamics of sheared colloidal suspensions~\cite{Pine2005, Corte2008,Wilken2020}.  In the simplest variant of RO, identical hyperspheres\footnote{Here we use the term `sphere' to refer to hyperspheres in arbitrary dimension.} (particles) are distributed randomly in $d$-dimensional space with a volume fraction $\phi$ \footnote{$\phi$ is the fraction of the d-dimensional space occupied by the spheres.}.  The model proceeds with the following dynamics: at each time step, isolated particles are left alone, while overlapping particles are considered ``active'', and are displaced in a random direction with a magnitude drawn uniformly from $[0,\epsilon]$.  These dynamics persist until either there are no active particles (an ``absorbing state''), or until the fraction of disks which are active reaches a steady state average value (an ``active state'').  

For a given $\epsilon$, there is a phase transition as a function of $\phi$: for $\phi < \phi_c$ a random initial state will ultimately arrive at an absorbing state, while for $\phi > \phi_c$ an absorbing state is never found, and the dynamics  continue forever.  The characteristic time to find an absorbing state or evolve to an active steady state diverges on either side of the transition, identifying this as a dynamical second order phase transition~\cite{henkel2008non}. However, unlike equilibrium transitions which have diverging critical fluctuations, the critical states of RO, although displaying no crystalline order, are characterized by severely depressed, hyperuniform~\cite{Torquato2003,Hexner2015} fluctuations in $d<4$, and non-divergent, gas-like fluctuations for $d\ge 4$. The universality class of RO has been shown to be the Manna class~\cite{Menon2009}.  

BRO modifies RO by adding a deterministic displacement along the line connecting the centers of overlapping particles (Fig.~\ref{critical densities}a).  
Specifically, if a pair of hyperspheres overlaps, each one is displaced by a vector $\bm{u} = \sqrt{\delta}\epsilon'\bm{\hat r} + \sqrt{(1-\delta)}\epsilon'\bm{\hat v}$, where $\delta$ is the proportion of the variance of displacements attributed to repulsive as opposed to random displacements, $\epsilon'$ is a random number drawn from  $[0, \epsilon]$, $\bm{\hat r}$ is the unit vector directed to the center of the displaced particle from the center of its overlapping neighbor, and $\bm{\hat v}$ is a random unit vector.
If $\delta = 0$,  we recover the original RO model. We previously found that, in $d=3$ in the $\epsilon \rightarrow 0$ limit, all $\delta>0$ approach the same densest critical point~\cite{Wilken2020RCP}. Therefore, for all simulations here we only consider $\delta = 1$. 

We first demonstrate that, as in $d=3$, BRO produces configurations with volume fractions that agree with the $d=4$ ($\phi_{RCP} \approx 0.46 \pm 0.005$) and $d=5$ ($\phi_{RCP} \approx 0.31 \pm 0.005$) values 
 in structures generated by other protocols~\cite{skoge2006packing}.
In Figure~\ref{critical densities}, we show the densities of the critical packings of identical hyperspheres in $d=1-5$. 
The densities of these critical packings depend on $\epsilon$ (the maximum displacement given to overlapping particles in BRO).   
In the limit $\epsilon \rightarrow 0$, we find $\phi_c \approx 1\pm0.001, \, 0.905\pm0.002, \, 0.640\pm0.001, \, 0.46\pm0.01$ and $0.30\pm0.01$ for $d = 1,2,3,4,5$ respectively.
In particular, BRO produces only ordered (crystalline) structures in $d=1$ and $2$, suggesting that RCP does not exist in $d=2$ unless polydispersity is added.

The values of $\phi_c$ seen in Fig.~\ref{critical densities}b suggest that a crystal results in both the trivial $d=1$ case and in $d=2$, where we obtain a triangular lattice of disks. 
Thus if our BRO$(\epsilon \rightarrow 0)$=RCP conjecture is correct, there are no random close packed structures which can be obtained in $d=2$ without either additional constraints on the simulation process or adding the frustration of polydispersity, an issue which has been a source of much recent speculation~\cite{atkinson2014existence,Zaccone2022, Blumenfeld2022, charbonneau2022comment, kranz2022comment, Chen2022comment}.
The density values for $d = 3-5$ are consistent with those obtained for RCP by other algorithms~\cite{skoge2006packing}.

\begin{figure}
  \centering
  \includegraphics[width=.9\linewidth]{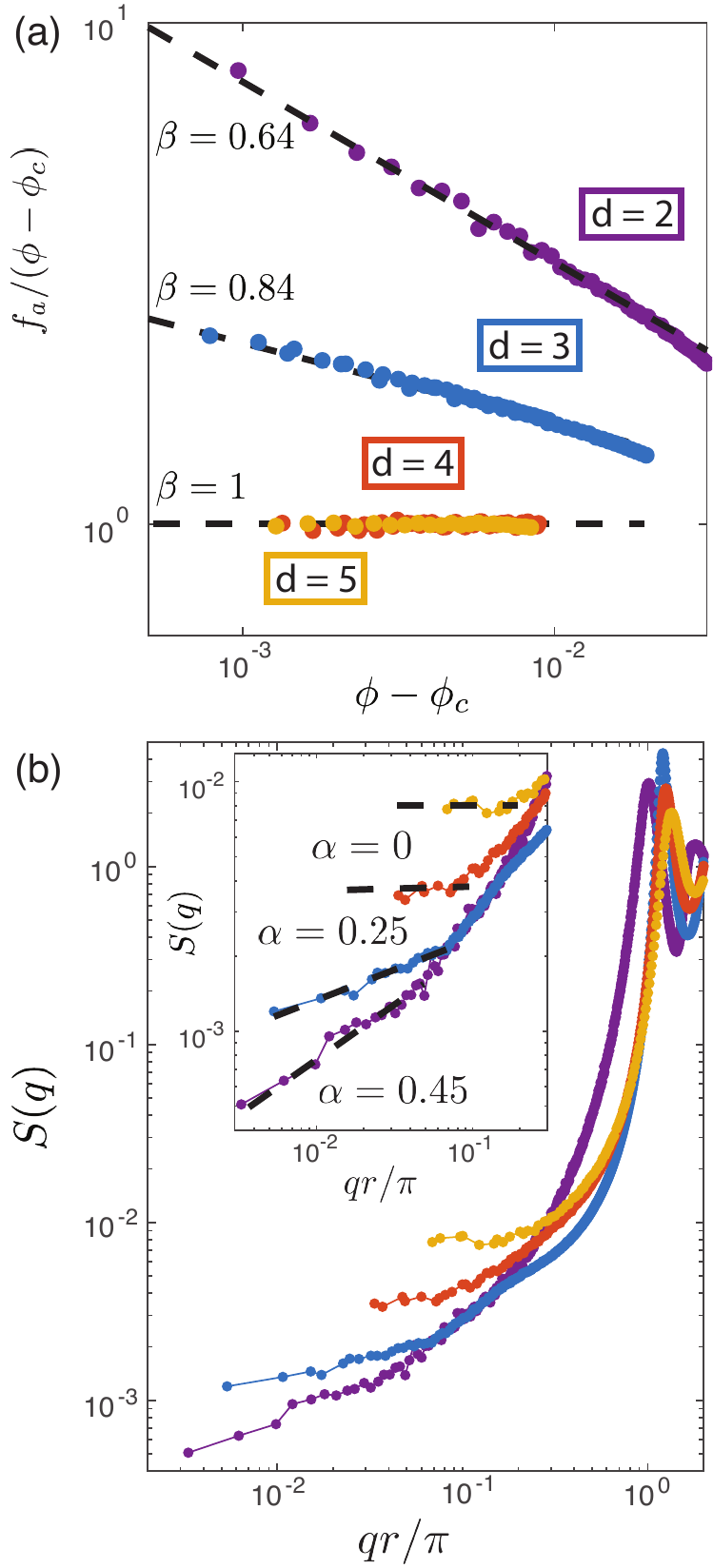}
  \caption{ {\bf BRO Critical Exponents.}
(a) Scaling of fraction of active particles, $f_a$, just above $\phi_c$ reveal Manna critical exponent $\beta$: $f_a \sim (\phi - \phi_c)^\beta$. For the Manna class, $\beta \approx 0.64$ for $d=2$, $\beta \approx 0.84$ for $d=3$, and $\beta = 1$ for $d \ge 4$. Plotted $f_a$ is normalized by $(\phi - \phi_c)$, the mean field behavior. Simulations in $d=2$ are composed of bidisperse disks with an equal number of particles with size ratio 1:$\sqrt{2}$. All simulations run with N=100,000 particles. Data shifted vertically for clarity. 
  (b) Structure factor of the densest critical states of BRO for $d = 2 - 5$. For all dimensions, the scaling of $S(q)$ follow the hyperscaling relation $\alpha = 2 - \eta_\perp$, where $S(q\rightarrow 0) \sim q^\alpha$. For $d=2$, (bidisperse) and $d=3$, the configurations are hyperuniform with $\alpha_{d=2} = 0.45$ and $\alpha_{d=3}=0.25$, but for $d=4$, 5, they are random not hyperuniform ($\alpha_{d=4-5} = 0$).
  }
  \label{SofQ}
\end{figure}

To verify that BRO is in the Manna universality class, we measure the steady-state activity $f_a$, the fraction of particles which are active, which is the order parameter for absorbing state systems. In our case, this quantity is the fraction of particles that overlap with at least one other particle. For the Manna class, we expect $f_a \sim (\phi-\phi_c)^\beta$, with $\beta \approx 0.64$ for $d=2$, $\beta \approx 0.84$ for $d=3$, and $\beta = 1$ for $d \ge 4$~\cite{henkel2008non}.
In Figure~\ref{SofQ}a, we plot the activity $f_a$ of BRO in $d = 2 - 5$ in the active phase $\phi - \phi_c>0$ near the critical point. For $d=2$, we use a bidisperse system to prevent crystallization; all other results are for monodisperse systems. For BRO in all these dimensions, we obtain critical behavior expected for the Manna class. In particular, for both $d = 4$ and 5, the activity shows the mean-field behavior $f_a \sim (\phi-\phi_c)$. 
This implies that the upper critical dimension is that of the Manna class, that is  $d_{uc}=4$, which is supported by our measurements of mean field exponents in both 4 and 5 dimensions. 

An indirect way of testing another critical exponent comes from the predicted scaling relation between the correlation function exponent, $\eta_\perp$,  and the hyperuniformity exponent measured in the static structure factor $S(q)$. For hyperuniform systems, $S(q \rightarrow 0) \sim q^\alpha$ and the prediction is $\alpha = 2-\eta_\perp$~\cite{Hexner2015}. For the Manna model, $\eta_\perp$ = 1.54, 1.74 for $d=2$ and $3$, and $\eta_\perp = 2$ is the mean field value appropriate for $d \ge 4$.
The structure factor calculations are carried out with the NFFT package and all $S(q)$ values are angularly averaged (i.e. $q = |{\bf q}|$). 
Configurations for $S(q)$ calculations consist of $10^6 \leq N \leq 5\times10^6$ spheres, and we average over 20 independent simulations.
 
For the bidisperse disk packings in $d=2$, as well as for the $d=3$ packings of identical spheres, we see that BRO yields critical packings in the $\epsilon \rightarrow 0$ limit which are hyperuniform, as seen by the small-$q$ power-law scaling of $S(q)$: $S(q) \sim q^\alpha$, with $\alpha_{d=2} \approx 0.45$ and $\alpha_{d=3} \approx 0.25$. 
These exponents were observed for other Manna class systems~\cite{Hexner2015} and the $d=3$ RCP data is the same as appears in~\cite{Wilken2020RCP}, where it is shown to agree with RCP configurations that result from both hard sphere compression and soft sphere minimization protocols. 
For $d=4$ and 5, we find $S(q \rightarrow 0) \sim $ {\it constant}, which implies $\alpha \approx 0$. These systems exhibit long range density fluctuations that are random (i.e. not hyperuniform).

\begin{figure}[t]
  \centering
  \includegraphics[width=.9\linewidth]{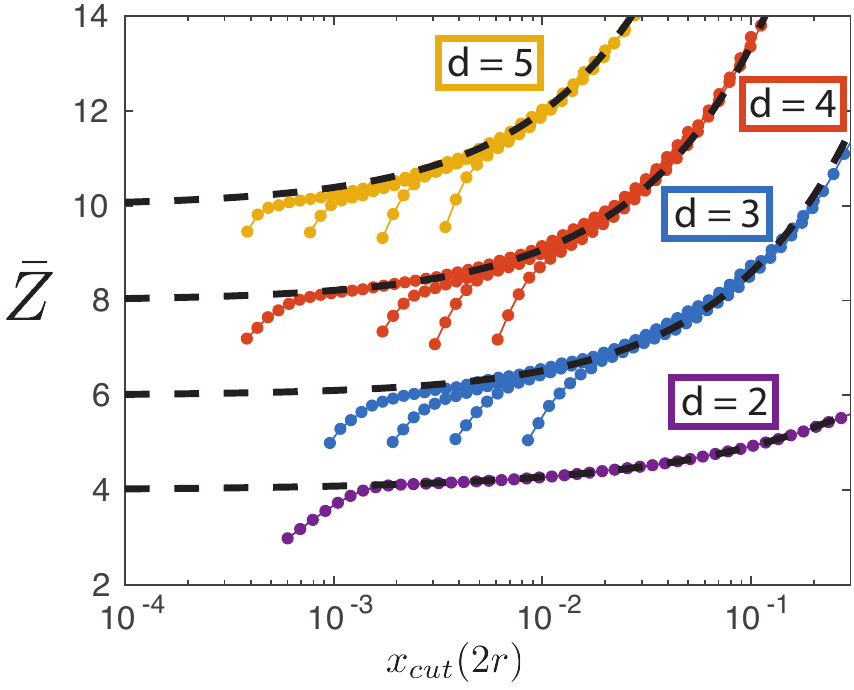}
  \caption{\textbf{Isostatic coordination of critical BRO.} Mean contact number, $\bar{Z}$, as a function of distance cutoff, $x_{cut}$, shows isostatic limiting behavior ($Z(x_{cut}, \epsilon \rightarrow 0) = 2d$).
  Dashed lines are power law fits to the form $\bar{Z} = Z_0 + A (x_{cut})^{0.7}$ for $4\epsilon<x_{cut}<0.1$ and $Z_0 = 2d \pm 0.05(2d)$ recovers the isostatic coordination.
  Contacts are undercounted when $x_{cut} \lesssim \epsilon$ because particle displacements are finite-positive and of order $\epsilon$ (traces are truncated for clarity). 
  For $d=3-5$, traces correspond to logarithmically spaced kick size ($\epsilon/r$ = $10^{-3} - 10^{-2}$). For $d=2$: $\epsilon/r = 10^{-3}$ and size distribution is bidisperse.
  All simulations: N=27,000 particles.
  }
  \label{Z}
\end{figure}
 
Having established that BRO is in the Manna universality class, we return to the question of whether it faithfully represents RCP. 
RCP states are isostatic, the number of contacts in the packings is equal to the number of degrees of freedom. 
In the absence of friction, this means that the average number of contacts per particle, $\bar{Z}$ equals 2d. 
In Figure~\ref{Z}, we plot the cumulative coordination $\bar{Z}(x_{cut})$, obtained by integrating the pair correlation function $g(r)$ between 0 and $x_{cut}$ for critical configurations of monodisperse particles in d = 3, 4, 5 and bidisperse particles in d = 2. 
For each dimension, data is shown for several values of $\epsilon/r$ between $10^{-2}$ and $10^{-3}$. 
We extrapolate to the limit $x_{cut}\rightarrow 0$ by fitting to the power law form $\bar{Z} = Z_0 + A (x_{cut})^{0.7}$~\cite{donev2005pair} for $4\epsilon<x_{cut}<0.1$ and recover $Z_0 = 2d \pm 0.05(2d)$.
This indicates that in the $\epsilon \rightarrow 0$ limit, BRO configurations are isostatic, as expected for Random Close Packing.

Another characteristic quantity related to $x_{cut}$ is the distribution of interparticle gaps near contact. 
An exact treatment~\cite{charbonneau2017glass} of hard sphere packings in infinite dimensions shows that the distribution $P(h)$ of near contacts or small gaps $h$ are distributed according to $P(h) \sim h^\gamma$, with $\gamma = 0.41269$.
In Figure~\ref{Gaps}b, we show the gap distribution for critical BRO configurations in $2\le d \le 5$ in the form of the pair correlation function $g(r)$, normalized by the $d = \infty$ case $g(r) \sim (r/\langle r \rangle - 1)^{-\gamma_{\infty}}$. 
For $d = 3$, we find that $\gamma = 0.5 \pm 0.03$, consistent with~\cite{silbert2002geometry}, while for d = 4, 5 we find $\gamma = 0.42 \pm 0.04$, consistent with the infinite dimensional case. 
This suggests that the mean field gap distribution exponent holds in $d \ge 4$, as would be expected for the Manna class $d_{uc}=4$.  

We note that the scaling of the gap distribution for BRO configurations seem to be at odds with previous work, which suggest that $\gamma = 0.41269$ holds in all dimensions $\geq$ 2~\cite{charbonneau2021finite,wang2022experimental}.
Previous investigations~\cite{charbonneau2012universal,lerner2013low} of the gap distribution removed undercoordinated particles i.e. ``rattlers" (defined as $Z<d+1$), because they do not contribute to the mechanical stress of the packing.
To investigate the role of rattlers in the gap distribution of BRO packings, we first identify the rattler fraction $f_r$ by fitting to an exponential form $f_r e^{-Ax_{cut}}$ (see Supplemental Figure S4) because BRO configurations do not contain true contacts.
For BRO configurations approaching critical $\langle Z \rangle = 2d$, we find a finite fraction of rattlers in $d=2,3$, consistent with previous estimates using other protocols (approximately $f_r \approx 0.05$ in $d=2$ and $f_r \approx 0.015-0.05$ in $d=3$)~\cite{lubachevsky1990geometric,donev2005pair,charbonneau2012universal,o2003jamming}.
In contrast, BRO rattler fractions appear to decay rapidly as the volume fraction approaches the critical point, therefore they have a negligible impact on the gap distribution for $d=4,5$ (Fig.~\ref{Gaps}).

We attribute the differences between our work and previous calculations of the gap distribution to the existence of a well-defined Manna class phase transition, and a well-defined $\phi_c$, and not to the presence of rattlers or model details (i.e. hard sphere compression vs. soft sphere energy minimization vs. BRO).
We calculate the gap distribution for all particles in the packing and compare to the distribution with rattlers removed, and we do not find a substantial difference in the gap distribution exponent $\gamma$ (Fig.~\ref{Gaps}b).
In $d=3$, the BRO gap distribution exponent agrees with RCP configurations produced by Lubachevsky-Stillinger hard sphere compression at the volume fraction corresponding to the minimum of an order parameter, even when rattlers are removed (see Supplemental Figure S3).
Additionally, we find that active states of BRO ($\phi>\phi_c$) have gap distribution exponents that agree with critical BRO in $d\ge 4$ and the infinite dimensional calculation (see Supplemental Figure S2).
The Manna class upper-critical dimensional behavior of the gap distribution only occurs for BRO configurations {\it at} the critical point, the gap distribution deviates from a power law form on the absorbing side of the transition while the gap exponent tends toward $\gamma_\infty$ on the active side.


We have shown that Biased Random Organization is a Manna universality class model that produces ensembles of Random Close Packed structures with non-crystalline, isostatic, isotropic, jammed and hyperuniform states as emergent properties rather than predefined outputs or constraints.
RCP simply emerges as the highest density critical point of BRO in arbitrary dimension with Manna class upper-critical dimension $d_{uc}=4$. 
As a well-defined mathematical model, BRO allows for a precise definition of RCP, displaying properties consistent with over 50 years of experimental, computational, and theoretical investigation. 
In the area of jamming, disordered and glassy materials, it presents a new and well-defined protocol for finding jammed states not only for hyperspheres but for arbitrarily shaped objects. 
Finally, results in $d=3$ show that adding global anisotropy to BRO, e.g. by shearing, results in an FCC crystal rather than RCP. 
Thus, studies of isotropic versus anisotropic BRO may provide new results on the age-old yet still important problem of the densest packing in higher dimensions and whether disordered structures ever pack more densely than ordered packings~\cite{torquato2006new}.

\begin{acknowledgments}
The authors thank Patrick Charbonneau, Aleks Donev, Rodrigo Guerra, Daniel Hexner,  Salvatore Torquato, Francesco Zamponi and Lilith Lulovitch for useful discussions. This work was supported by the National Science Foundation Physics of Living Systems Grant No. 1504867 and U.S. Department of Energy, Office of Science, Office of Basic Energy Sciences under Award Number DE-SC-0020976 (A.Z.G.) for simulations, NSF CBET 1832260: GOALI, and DOE DE-SC0007991 (P.C.) for theory and calculations, and the MRSEC Program of the National Science Foundation under Grant No. DMR-1420073 (S.W.) for comparison simulations. D.L. acknowledges the support of the US-Israel Binational Science Foundation (Grant No. 2014713) and the Israel Science Foundation (Grant No. 1866/16). 
\end{acknowledgments}

%

\end{document}